\documentclass[twocolumn,showpacs,preprintnumbers,amsmath,amssymb]{revtex4}
\usepackage{graphicx}% Include figure files
\usepackage{dcolumn}% Align table columns on decimal point
\usepackage{bm}% bold math

\def\dslash{/\kern-.1em \partial}

\def\deter{\hbox{det}}

\def\al{\alpha}
\def\be{\beta}
\def\ga{\gamma}

\def\la{\lambda}

\def\si{\sigma}

\def\ta{\tau}

\def\ph{\phi}

\def\ps{\psi}

\def\Ga{\Gamma}

\def\mn{{\mu\nu}}

\def\cl{{\cal L}}

\def\fr#1#2{{{#1} \over {#2}}}

\def\half{{\textstyle{1\over 2}}}
\def\frac#1#2{{\textstyle{{#1}\over {#2}}}}

\def\lsim{\mathrel{\rlap{\lower4pt\hbox{\hskip1pt$\sim$}}
    \raise1pt\hbox{$<$}}}
\def\gsim{\mathrel{\rlap{\lower4pt\hbox{\hskip1pt$\sim$}}
    \raise1pt\hbox{$>$}}}
\def\sqr#1#2{{\vcenter{\vbox{\hrule height.#2pt
         \hbox{\vrule width.#2pt height#1pt \kern#1pt
         \vrule width.#2pt}
         \hrule height.#2pt}}}}

\def\lrDmu{\stackrel{\leftrightarrow}{D_\mu}}

\def\lrDnu{\stackrel{\leftrightarrow}{D^\nu}}

\newcommand{\beq}{\begin{equation}}
\newcommand{\eeq}{\end{equation}}
\newcommand{\bea}{\begin{eqnarray}}
\newcommand{\eea}{\end{eqnarray}}

\begin{document}

\preprint{NCF/002}

\title{One-Loop Renormalization of the Electroweak Sector with Lorentz Violation}% Force line breaks with \\

\author{Don Colladay}
\email{colladay@ncf.edu}
\author{Patrick McDonald}%
 \email{mcdonald@ncf.edu}
\affiliation{%
New College of Florida\\ Sarasota, FL, 34243, U.S.A.
}%

\date{\today}% It is always \today, today,

\begin{abstract}
The one-loop renormalizability of the electroweak sector of the
Standard Model Extension with Lorentz violation is studied.
Functional determinants are used to calculate the one-loop contributions
of the higgs, gauge bosons and fermions to the one-loop effective action.
The results are consistent with multiplicative renormalization of the
SME coupling constants.
Conventional Electroweak symmetry breaking is effectively unaltered relative to
the standard case as the renormalized SME parameters are sufficient to absorb
all infinite contributions.

\end{abstract}

\pacs{Valid PACS appear here}% PACS, the Physics and Astronomy
\maketitle

\section{\label{sec:intro}Introduction}

The Standard Model Extension (SME) provides a framework within the context
of conventional quantum field theory that allows for general Lorentz-breaking
effects.
The construction generates all possible couplings of standard model fields to
constant background fields that serve as a source for spontaneous Lorentz breaking.

In a recent series of articles, various one-loop renormalization calculations
have been performed.
All of these have indicated that multiplicative renormalization still works, even when
Lorentz symmetry is not intact.
Explicit verification in the case of Lorentz-breaking theories is important since
conventional techniques frequently make use of Lorentz invariance
arguments to establish renormalizbility.
This paper extends these previously obtained results to the electroweak sector of the SME.
More specifically, the functional determinant formalism is used to
calculate radiative corrections to one-loop in the Higgs, fermion, and gauge boson
sectors of the electroweak model.
The renormalization is incorporated using multiplicative factors as in the
conventional case.
This renormalization procedure is formally carried out before $SU(2)\times U(1)$ breaking
so that the symmetries can be utilized fully.

The investigation of the renormalizability properties of the SME
was first started in \cite{klp1,klp2} where the authors
studied one-loop radiative corrections for QED with Lorentz
violation.  In this prior work, the one-loop
renormalizability of general Lorentz and CPT violating QED was established.
The manuscript \cite{klp1} includes an analysis of the explicit one-loop
structure of Lorentz-violating QED and the resulting running of the
couplings.  The authors established that conventional multiplicative
renormalization succeeds and they find that the beta functions indicate a variety of
running behaviors, all controlled by the running of the charge.
Portions of this analysis have been extended to allow for a
curved-space background \cite{bps}, while other analysis involved
finite, but undetermined radiative corrections due to CPT violation
\cite{kafstuff1,kafstuff2,kafstuff3,kafstuff4,kafstuff5,kafstuff6,kafstuff7}.
In addition, recent papers have addressed anomalies in the presence of Lorentz-violating
terms \cite{anom1,anom2}.
The main results indicate that the anomaly is present even in the absence of Lorentz
symmetry and the fundamental nature of the anomaly is essentially the same as in the conventional
case.

The Lorentz violating QED results of \cite{klp1} were extended to non-abelian gauge
theories including QCD in \cite{cm1,cm2} where the authors established that
Yang-Mills theory is renormalizable at one-loop, provided the gauge group remains
unbroken.
The electroweak sector presents additional challenges, mainly due to the $SU(2)\times U(1)$
symmetry breaking and the parity violating fermion sectors.
In addition, the Higgs participates as a scalar field that was not considered
in previous work on the subject.
The present paper focuses on the functional determinants that integrate out the
Higgs and fermion sectors as the gauge sector
has already been handled in sufficient detail in \cite{cm2}.

The current work should be viewed as part of an extensive, systematic
investigation of Lorentz violation and its possible implications for
Planck-scale physics \cite{kps1,kps2,kps3,kps4,kps5,kps6,kps7,kps8,kps9}.
Recent work involving Lorentz violation and cosmic microwave background fluctuations
\cite{kmewes} suggest that the SME might play a useful role in cosmology.  In
addition to the above, the SME formalism has been extended to include
gravity \cite{alangrav1,alangrav2,alangrav3}, where it has been suggested that Lorentz
violation provides an alternative means of generating General
Relativity \cite{kpa,kpb}.

Some other work relevant to the current paper includes a study of deformed
instantons in pure Yang-Mills theory with Lorentz Violation \cite{cmjmp1,cmjmp2},
an analysis of the Casimir effect in the presence of Lorentz violation
\cite{ft}, an analysis of gauge invariance of Lorentz-violating
QED at higher-orders \cite{ba}, and possible effects due to nonpolynomial interactions \cite{kalt}.
Some investigations into possible
Lorentz-violation induced from the ghost sector of
scalar QED have also been performed \cite{altghost}.
Higher powers of spatial derivatives that violate Lorentz invariance
have been used to argue improved behavior of renormalization for scalar and gauge theories
\cite{ans1,ans2} at higher-order.
In addition, functional determinants have been used to compute finite corrections
to CPT-violating gauge terms arising from fermion violation \cite{funcdet}.

\section{\label{sec:nandc}Notation and Conventions}

In this paper we adopt the conventions used in \cite{ck1,ck2} to
define standard model fields and Lorentz-violating couplings.
The electroweak sector contains left- and right-handed lepton and quark
multiplets denoted as

$$
L_A = \left( \begin{array}{c} \nu_A \\ l_A
\end{array} \right)_L
\quad , \quad
R_A = (l_A)_R
\quad ,
$$
\beq
Q_A = \left( \begin{array}{c} u_A \\
d_A \end{array} \right)_L ~~ , ~~
U_A = (u_A)_R ~~ , ~~ D_A = (d_A)_R
\quad ,
\eeq
where
\beq
\ps_L \equiv \frac 1 2 ( 1 - \ga_5 ) \ps
\quad , \qquad
\ps_R \equiv \frac 1 2 ( 1 + \ga_5 ) \ps
\quad ,
\label{handproj}
\eeq
as usual,
and where $A = 1,2,3$ labels the flavor:
$l_A \equiv (e, \mu, \ta)$,
$\nu_A \equiv (\nu_e, \nu_\mu, \nu_\ta)$,
$u_A \equiv (u,c,t)$, $d_A \equiv (d,s,b)$.
We denote the Higgs doublet by $\ph$.

The lagrangian terms in the
usual SU(2) $\times$ U(1) electroweak sector of the
minimal standard model are
\beq
\cl_{\rm lepton} =
\half i \overline{L}_A \ga^{\mu} \lrDmu L_A
+ \half i \overline{R}_A \ga^{\mu} \lrDmu R_A
\label{smlepton}
\quad ,
\eeq
\bea
\cl_{\rm quark} &=&
\half i \overline{Q}_A \ga^{\mu} \lrDmu Q_A
+ \half i \overline{U}_A \ga^{\mu} \lrDmu U_A
\nonumber \\ &&
+ \half i \overline{D}_A \ga^{\mu} \lrDmu D_A
\quad ,
\label{smquark}
\eea
\bea
\cl_{\rm Yukawa} &=&
- \left[ (G_L)_{AB} \overline{L}_A \ph R_B
+ (G_U)_{AB} \overline{Q}_A \ph^c U_B
\right.
\nonumber \\ &&
\left.
\qquad + (G_D)_{AB} \overline{Q}_A \ph D_B
\right]
+ {\rm h.c.}
\quad ,
\label{smyukawa}
\eea
where h.c. indicates the hermitian conjugate,
\beq
\cl_{\rm Higgs} =
(D_\mu\ph)^\dagger D^\mu\ph
+\mu^2 \ph^\dagger\ph - \fr \la {3!} (\ph^\dagger\ph)^2
\quad ,
\label{smhiggs}
\eeq
\beq
\cl_{\rm gauge} =
-\half {\rm Tr} (W_{\mu\nu}W^{\mu\nu})
-\frac 1 4 B_{\mu\nu}B^{\mu\nu} ~ .
\label{smgauge}
\eeq
The Yukawa coupling $G_L$ for the leptons may be diagonalized in the usual way
before any $SU(2) \times U(1)$ symmetry breaking occurs.
The quark couplings $G_U$ and $G_D$ are more complicated due to the existence of right-handed
up quarks.
It is only possible to simultaneously diagonalize these couplings in a specific
gauge, so they are left arbitrary at this point.
Note that massive neutrinos can be incorporated easily by making the structure of
the neutrino sector match that of the quark sector.
This is not done here so that the two sectors may be contrasted more effectively.

The corresponding Lorentz-violating terms in the SME are given for each sector as they
appear in the following sections of the paper.

\section{\label{sec:func}Higgs sector corrections}

In this section we introduce the Lorentz-violating terms in the SME involving Higgs couplings.
Recall, the one-loop effective action for a field theory can be written as a functional
integral over fields $\Psi$:
\beq
\exp{i\Ga[\Psi]} = \int{\mathcal D}\Psi e^{i\int d^4 x {\mathcal
    L}[\Psi]}~.
\eeq
The effective action is constructed by writing the underlying fields as the sum
of classical background fields and fluctuating quantum fields.
The effective action is given by
a classical term perturbed by terms quadratic in the fluctuation.  The
quadratic term gives rise to a Gaussian integral, which in turn can
be described by a functional determinant \cite{ps}.  Using
${\mathcal L}_{cl} = {\mathcal L}_0 + {\mathcal L}_{c.t.}$
for the classical Lagrangian as a function of
the background field where ${\mathcal L}_{c.t.}$  is the counterterm Lagrangian,
the expression generates terms of the form
\begin{equation}\label{expldet3.1}
\exp{i\Ga[\Psi]} = e^{i\int d^4x {\mathcal L}_{cl}}
   \deter(\Delta)^{n}~,
\end{equation}
where the $\Delta$ are operators which are given explicitly below, and $n$
is an exponent that depends on the field type.
To compute the above determinants, dimensional regularization is used.
Each determinant is treated separately, beginning with the pure Yang-Mills
gauge field contribution.  The calculation is performed to first order in Lorentz violating
parameters.  As this is the case, the computations of the various terms decouple
and the CPT-even and CPT-odd cases can be treated independently.

The conventional Higg's sector lagrangian with $SU(2)_L$ doublet $\phi$ is written as
\beq
{\cal L}_{\rm Higgs} = (D_\mu \phi)^\dagger D^\mu \phi + \mu^2 \phi^\dagger \phi - {\lambda \over 3!}(\phi^\dagger \phi)^2 ~.
\eeq
The covariant derivative acting on $\phi$ is
\beq
D_\mu \phi = \left( \partial_\mu - i \frac g 2 W_\mu^a \sigma^a - i \frac {g^\prime} {2} B_\mu\right) \phi~.
\eeq
For notational simplicity, it is convenient to introduce the quantity
\beq
A_\mu = \frac g 2 W_\mu^a \sigma^a + \frac{g^\prime} {2} B_\mu ~.
\eeq
The Lorentz-violating contributions split into CPT even and CPT odd terms
\bea
{\cal L}_{\rm Higgs}^{\rm CPT-even} = & & \half (k_{\phi\phi})^\mn (D_\mu \phi)^\dagger D_\nu \phi + h.c. \nonumber \\
& & -\half (k_{\phi B})^\mn \phi^\dagger \phi B_\mn - \half (k_{\phi W})^\mn \phi^\dagger W^\mn \phi ~, \nonumber \\
\eea
and
\beq
{\cal L}_{\rm Higgs}^{\rm CPT-odd} = i (k_\phi)^\mu \phi^\dagger D^\mu \phi + h.c.
\eeq

We will write the Higgs fields as the sum of a classical background
field (denoted with an underline) and a fluctuating quantum field:
\bea
\phi & \rightarrow & \underline{\phi} +{\phi}.
\eea
First, the one-loop contribution of the term $k_{\phi\phi}$ is calculated.
The quadratic contribution to the lagrangian is
\beq
{\mathcal L}_{higgs} = \phi^\dagger \left[ -D^2 + \mu^2 - \half \left[(k_{\phi\phi})^\mn D_\mu D_\nu + h.c.\right] \right] \phi
\eeq
Integration over the Higgs fields yields logdet of the operator in the brackets.
To facilitate calculation, $k_{\phi\phi} = k_{\phi\phi}^R + i k_{\phi\phi}^I$ is split into
a symmetric real part and an antisymmetric imaginary piece.
The kinetic, field-independent piece of the operator
\beq
P = -(\eta^\mn + k_{\phi\phi}^\mn) \partial_\mu \partial_\nu + \mu^2~,
\eeq
is factored out of the expression.
The inverse of this operator is written using the Fourier expansion
\beq
P^{-1} = \int {d^4 p ~ e^{-i p\cdot (x-y)}\over (2 \pi)^4 \left((\eta^\mn + k_{\phi\phi}^\mn)p_\mu p_\nu + \mu^2\right)}~.
\eeq
Note that $\mu^2$ appears with opposite sign to the usual scalar field propagator mass term.
This occurs because
the renormalization is being performed {\it before} spontaneous breaking of $SU(2)\times U(1)$ is implemented.
The operator expansion
\beq
{1 \over A + B} \simeq {1 \over A} - {1 \over A}B{1 \over A} + \cdots~,
\eeq
is then used to expand the inverse kinetic operator for small $k_{\phi\phi}$.
The result of a calculation involving the real component yields the Lorentz-violating contribution
\bea
logdet \left[ -D^2 - \half (k_{\phi\phi}^R)^\mn D_\mu D_\nu \right] = \nonumber \\
{1 \over 6} {i \over (4 \pi)^2}
 \Gamma(2 -\frac d 2) (k_{\phi\phi}^R)^\mn Q^\mn
\eea
where the gauge invariant operator $Q^\mn$ is defined as
\begin{equation}\label{qmunu}
Q^{\mu\nu} = tr \int\frac{d^4k}{(2\pi)^4} (k^\mu k^\nu{\underline
  A}^2 -2 k^\mu \underline A^\nu k\cdot {\underline A} + k^2 {\underline
  A}^\mu{\underline A}^\nu)~.
\end{equation}
Note that this term contributes radiative corrections to the anti-self-dual components of
the Lorentz-violating gauge terms
\bea
{\cal L}_{\rm gauge}^{\rm CPT-even} & = & - {1 \over 2} Tr (k_W)^{\mu\nu\alpha\beta} W_{\mu\nu} W_{\al\be}
\nonumber \\ &&
- {1 \over 4} (k_B)^{\mu\nu\alpha\beta} B_{\mn}B_{\al\be}.
\label{gauge}
\eea
A calculation using the antisymmetric component of $k_{\phi\phi}$ yields
\bea
logdet[-D^2 -\half (k_{\phi\phi}^I)^\mn D_\mu D_\nu] = \nonumber \\
-{g^\prime \over 4} \int {d^4 p \over (2 \pi)^4 (p^2 + \mu^2)} (k_{\phi\phi}^I)^\mn B_{\mn}
\eea
a quadratically divergent linear field instability.
This suggests that $k_{\phi\phi}^I$ should be set to zero in any sensible theory,
although it is possible to arrange a cancellation as described next.

The remaining CPT-even Higgs corrections are the $k_{\phi W}$ and $k_{\phi B}$
couplings.
The contribution of $k_{\phi W}$ is zero due to the Lie Algebra trace, but the
term $k_{\phi B}$ contributes
\bea
logdet[-D^2 -\half (k_{\phi B})^\mn B_{\mn}] = \nonumber \\
-{1 \over 2} \int {d^4 p \over (2 \pi)^4 (p^2 + \mu^2)} (k_{\phi B})^\mn B_{\mn}
\eea
again introducing a linear instability.
If the bare coefficients are chosen appropriately, it is possible to obtain
a cancellation $\frac {g^\prime} {2} k_{\phi\phi}^I + k_{\phi B} = 0$
at one loop.
This choice simply corresponds to choosing the $k_{\phi B}$ term to cancel
the corresponding $k_{\phi \phi}^I$ coupling to the $B$ field in the original
lagrangian.
In general, there is no such restriction and a quadratically divergent
linear instability in $B_{\mn}$ will result.

The remaining CPT-odd Higgs sector term is the vector $k_\phi$.
It's contribution to the divergent piece of the determinant is
in fact zero as the various terms cancel out in the equations.
In fact, the field redefinition
\beq
\phi \rightarrow e^{i k_\phi \cdot x} \phi~,
\eeq
of the Higgs field eliminates this term
entirely from the theory to all orders,
contributing only to the mass parameter for the Higgs at second order
in $k_\phi$.
In addition, this redefinition removes the nontrivial vacuum expectation value found
for the $Z$ boson in \cite{ck2}.
This implies that the electroweak symmetry breaking is unaltered from
the conventional case when the fields are appropriately defined.

\section{\label{sec:func}Fermion sector corrections}

The left- and right-handed fermion fields are treated differently
with respect to the covariant derivative, therefore, they must be
separated in the construction of the determinant.
To accomplish this in the leptonic sector,
the fermion fields in each generation are arranged into a six-component
multiplet of the form
\beq
L_A =
\left(
\begin{array}{c}
\nu_L \\ l_L \\ \l_R
\end{array}
\right)_A.
\eeq
The Lorentz-invariant portion of the Lagrangian can be written in the form
${\cal L} = \overline L M L$ using the cross-generational matrix
\beq
M_{AB} =
\left(
\begin{array}{ccc}
{g \over 2} \not \overline W^3 - {g^\prime \over 2} \not \overline B & {g \over \sqrt{2}}\not \overline W^+ &
(G_L)_{AB} \phi^+ \\
{g \over \sqrt{2}}\not \overline W^- & - {g \over 2} \not \overline W^3 - {g^\prime \over 2} \not \overline B &
(G_L)_{AB} \phi^0 \\
(G_L^\dagger)_{AB} \phi^- & (G_L^\dagger)_{AB} \phi^{*0} & -g^\prime \not B
\end{array}
\right),
\eeq
Where the Feynman slash notation indicates contraction with one of the $2\times 2$
representations $\si^\mu = (Id, \vec \sigma)$, or $\overline \si^\mu = (Id, - \vec \sigma)$ for the right- and left-handed
fields respectively:
\beq
\not A = \si^\mu A_\mu  ,
\quad
\not \overline A = \overline \si^\mu A_\mu .
\eeq
The Lorentz-violating terms can be easily included into the above
matrix.
In the $4 \times 4$ Chiral representation for $\gamma^\mu$, they take the form
\bea
\cl^{\rm CPT-even}_{\rm lepton} &=&
\half i (c_L)_{\mu\nu AB} \overline{L}_A \ga^{\mu} \lrDnu L_B
\nonumber\\ &&
+ \half i (c_R)_{\mu\nu AB} \overline{R}_A \ga^{\mu} \lrDnu R_B
\label{lorvioll}
\quad ,
\eea
\beq
\cl^{\rm CPT-odd}_{\rm lepton} =
- (a_L)_{\mu AB} \overline{L}_A \ga^{\mu} L_B
- (a_R)_{\mu AB} \overline{R}_A \ga^{\mu} R_B
\label{cptvioll}
\quad ,
\eeq
and the Yukawa terms are
\beq
\cl^{\rm CPT-even}_{\rm Yukawa} =
- \half
\left[
(H_L)_{\mu\nu AB} \overline{L}_A \ph \si^{\mu\nu} R_B
+ {\rm h.c.} \right].
\label{loryukawa}
\eeq
The kinetic, field-independent portion of the operator can be written
as
\beq
P_{AB} =
\left(
\begin{array}{ccc}
(P_L)_{AB} & 0&0 \\
0 & (P_L)_{AB} &0 \\
0 & 0 & (P_R)_{AB}
\end{array}
\right),
\eeq
where $(P_L)_{AB} = i (\overline \not \partial \delta_{AB} + (c_L)_{\mn AB} \overline \sigma^\mu \partial^\nu)$,
and $(P_R)_{AB} =  i (\not\partial \delta_{AB} + (c_R)_{\mn AB} \sigma^\mu \partial^\nu) $.
The contribution of the $c$-terms to the logdet in this case yields $T^{\rm lep} = T_B^{\rm lep} + T_W$ where
\beq
T_B^{\rm lep} = - {1 \over 3} {i \over (4 \pi)^2}g^{\prime 2} \Ga (2 - \frac d 2)Tr(c_L + 2c_R)_{\mn}Q_B^{\mn},
\eeq
and
\beq
T_W = - {1 \over 3} {i \over (4 \pi)^2}g^2 \Ga (2 - \frac d 2)Tr(c_L)_{\mn}Q_W^{\mn},
\label{twcalc}
\eeq
where $Q_B^\mn$ and $Q_W^\mn$ are defined as in eq.(\ref{qmunu}) with $\underline A$
replaced with $B$ or $W$ respectively.
The trace includes a summation over generational indices that have been
suppressed for notational simplicity.
This contribution can be absorbed into the Lorentz-violating gauge $k_B$ and $k_W$ terms
given in eq.(\ref{gauge}).
The Lorentz-violating Yukawa terms yield traces over sigma matrices that vanish, so they don't
contribute to one-loop radiative corrections.
In addition, the CPT-odd leptonic terms can be eliminated using field redefinitions, so they don't
contribute either.

The quark sector may be handled similarly by defining the multiplet
\beq
L_A =
\left(
\begin{array}{c}
u_L \\ d_L \\ u_R \\ d_R
\end{array}
\right)_A.
\eeq
The calculation proceeds in exactly the same manner as in the leptonic sector
with $M_{AB}$ replaced by the corrseponding $4\times4$ matrix.
As in the leptonic sector, the only contribution arises from the $c$
couplings
\bea
\cl^{\rm CPT-even}_{\rm quark} &=&
\half i (c_Q)_{\mu\nu AB} \overline{Q}_A \ga^{\mu} \lrDnu Q_B
\nonumber\\ &&
+ \half i (c_U)_{\mu\nu AB} \overline{U}_A \ga^{\mu} \lrDnu U_B
\nonumber \\ &&
+ \half i (c_D)_{\mu\nu AB} \overline{D}_A \ga^{\mu} \lrDnu D_B
\label{lorvioll}
\quad ,
\eea
with the result for logdet of $T^q = T_B^q + T_W$ with $T_W$ the same as in the leptonic
calculation of eq.(\ref{twcalc}) and
\beq
T_B^q = - {1 \over 27} {i \over (4 \pi)^2}g^{\prime 2} \Ga (2 - \frac d 2)Tr(c_Q + 8c_U + 2c_D)_{\mn}Q_B^{\mn}.
\eeq
The difference in coupling factors are due to the standard hypercharge assignments of the quark fields
relative to the leptonic fields.

\section{\label{sec:sum}Electroweak Symmetry Breaking}

Once the coupling constants have been renormalized, it is a simple matter to incorporate
the electroweak symmetry breaking.
Specifically, a term can be added to the Lagrangian of the form
\beq
\cl_{SB} = {\bf s} \cdot \phi(x),
\eeq
where ${\bf s}$ is a small, external source field that induces the breaking.
The vacuum expectation value for the Higgs field is therefore fixed to point
along the direction of ${\bf s}$.
The inverse of the renormalization factor used to rescale $\phi$ can be used to renormalize the
external source to keep the symmetry breaking term finite.
This procedure is the same as the procedure used to incorporate renormalization in the linear
sigma model with symmetry breaking \cite{iz}.

Once the renormalization factors are included in the standard SME parameters and they are
rendered finite, it is a simple matter to perform
an extremization of the static electroweak potential.
This calculation has already been performed in \cite{ck2} and will not be reproduced
here.
One additional feature worth mentioning is that the diagonalization of the mass
matrices for the photon and $Z$ introduces quadratic photon-Z couplings proportional
to the difference between the CPT-even gauge couplings $k_W$ and $k_B$.
This may lead to interesting novel experimental effects and may place very stringent
bounds on the difference between these two parameters.

\section{\label{sec:sum}Summary}

Functional determinant techniques can be easily adapted to the electroweak sector
by performing the renormalization prior to $SU(2)\times U(1)$ breaking.
It is found that conventional multiplicative renormalization factors suffice to
renormalize the theory to one loop as with the other sectors of the SME.
Electroweak breaking is then essentially the same as in the usual standard model,
provided the proper field redefinition is implemented on the phase of the Higgs field.

This paper (together with references \cite{cm1,cm2}) exhausts the uses of functional determinants
in computing one-loop renormalization effects in the SME.
This serves as an important first step towards future work that will hopefully involve demonstrating
full renormalization to all orders in perturbation theory for the SME.
The one-loop renormalizability of the SME is promising as it is used effectively in the
standard case to argue all-orders results, but in the standard case, Lorentz invariance
is frequently used in the arguments, so the details of the argument in the absence of
Lorentz invariance would be very interesting.

One interesting fact that results from the analysis is that parameters that are not
bounded very tightly in the tree-level theory (such as tau couplings...) contribute
to the one-loop effective action suppressed by the square of the appropriate coupling.
This indicates that very stringent photon bounds (for example) may be used to estimate
reasonable levels for other coupling constants in the SME that are so far unbounded
by direct experiment.

\begin{acknowledgments}
We wish to acknowledge the support of New College of Florida's faculty development
funds that contributed to the successful completion of this project.
\end{acknowledgments}

%\newpage %Just because of unusual number of tables stacked at end
\bibliography{higgsrenorm}% Produces the bibliography via BibTeX.

\begin{thebibliography}{42}
\expandafter\ifx\csname natexlab\endcsname\relax\def\natexlab#1{#1}\fi
\expandafter\ifx\csname bibnamefont\endcsname\relax
  \def\bibnamefont#1{#1}\fi
\expandafter\ifx\csname bibfnamefont\endcsname\relax
  \def\bibfnamefont#1{#1}\fi
\expandafter\ifx\csname citenamefont\endcsname\relax
  \def\citenamefont#1{#1}\fi
\expandafter\ifx\csname url\endcsname\relax
  \def\url#1{\texttt{#1}}\fi
\expandafter\ifx\csname urlprefix\endcsname\relax\def\urlprefix{URL }\fi
\providecommand{\bibinfo}[2]{#2}
\providecommand{\eprint}[2][]{\url{#2}}

\bibitem[{\citenamefont{Kostelecky et~al.}(2002)\citenamefont{Kostelecky, Lane,
  and Pickering}}]{klp1}
\bibinfo{author}{\bibfnamefont{V.~A.} \bibnamefont{Kostelecky}},
  \bibinfo{author}{\bibfnamefont{C.}~\bibnamefont{Lane}}, \bibnamefont{and}
  \bibinfo{author}{\bibfnamefont{A.}~\bibnamefont{Pickering}},
  \bibinfo{journal}{Phys.\ Rev.\ D} \textbf{\bibinfo{volume}{65}},
  \bibinfo{pages}{056006} (\bibinfo{year}{2002}).

\bibitem[{\citenamefont{Kostelecky and Pickering}(2003)}]{klp2}
\bibinfo{author}{\bibfnamefont{V.~A.} \bibnamefont{Kostelecky}}
  \bibnamefont{and}
  \bibinfo{author}{\bibfnamefont{A.}~\bibnamefont{Pickering}},
  \bibinfo{journal}{Phys.\ Rev.\ Lett.} \textbf{\bibinfo{volume}{91}},
  \bibinfo{pages}{031801} (\bibinfo{year}{2003}).

\bibitem[{\citenamefont{Berredo-Peixoto and Shapiro}(2006)}]{bps}
\bibinfo{author}{\bibfnamefont{G.}~\bibnamefont{Berredo-Peixoto}}
  \bibnamefont{and} \bibinfo{author}{\bibfnamefont{I.}~\bibnamefont{Shapiro}},
  \bibinfo{journal}{Phys.\ Lett.\ B} \textbf{\bibinfo{volume}{642}},
  \bibinfo{pages}{153} (\bibinfo{year}{2006}).

\bibitem[{\citenamefont{Carroll et~al.}(1990)\citenamefont{Carroll, Field, and
  Jackiw}}]{kafstuff1}
\bibinfo{author}{\bibfnamefont{S.}~\bibnamefont{Carroll}},
  \bibinfo{author}{\bibfnamefont{G.}~\bibnamefont{Field}}, \bibnamefont{and}
  \bibinfo{author}{\bibfnamefont{R.}~\bibnamefont{Jackiw}},
  \bibinfo{journal}{Phys.\ Rev.\ D} \textbf{\bibinfo{volume}{41}},
  \bibinfo{pages}{1231} (\bibinfo{year}{1990}).

\bibitem[{\citenamefont{Jackiw and Kostelecky}(1999)}]{kafstuff2}
\bibinfo{author}{\bibfnamefont{R.}~\bibnamefont{Jackiw}} \bibnamefont{and}
  \bibinfo{author}{\bibfnamefont{A.}~\bibnamefont{Kostelecky}},
  \bibinfo{journal}{Phys.\ Rev.\ Lett.} \textbf{\bibinfo{volume}{82}},
  \bibinfo{pages}{3572} (\bibinfo{year}{1999}).

\bibitem[{\citenamefont{Perez-Victoria}(1999)}]{kafstuff3}
\bibinfo{author}{\bibfnamefont{M.}~\bibnamefont{Perez-Victoria}},
  \bibinfo{journal}{Phys.\ Rev.\ Lett.} \textbf{\bibinfo{volume}{83}},
  \bibinfo{pages}{2518} (\bibinfo{year}{1999}).

\bibitem[{\citenamefont{Chen}(1999)}]{kafstuff4}
\bibinfo{author}{\bibfnamefont{W.}~\bibnamefont{Chen}},
  \bibinfo{journal}{Phys.\ Rev.\ D} \textbf{\bibinfo{volume}{60}},
  \bibinfo{pages}{085007} (\bibinfo{year}{1999}).

\bibitem[{\citenamefont{Chung}(1999)}]{kafstuff5}
\bibinfo{author}{\bibfnamefont{J.}~\bibnamefont{Chung}},
  \bibinfo{journal}{Phys.\ Rev.\ D} \textbf{\bibinfo{volume}{60}},
  \bibinfo{pages}{127901} (\bibinfo{year}{1999}).

\bibitem[{\citenamefont{Altschul}(2004{\natexlab{a}})}]{kafstuff6}
\bibinfo{author}{\bibfnamefont{B.}~\bibnamefont{Altschul}},
  \bibinfo{journal}{Phys.\ Rev.\ D} \textbf{\bibinfo{volume}{69}},
  \bibinfo{pages}{125009} (\bibinfo{year}{2004}{\natexlab{a}}).

\bibitem[{\citenamefont{Altschul}(2004{\natexlab{b}})}]{kafstuff7}
\bibinfo{author}{\bibfnamefont{B.}~\bibnamefont{Altschul}},
  \bibinfo{journal}{Phys.\ Rev.\ D} \textbf{\bibinfo{volume}{70}},
  \bibinfo{pages}{101701} (\bibinfo{year}{2004}{\natexlab{b}}).

\bibitem[{\citenamefont{Arias et~al.}(2007)\citenamefont{Arias, H.Falomir,
  Gamboa, Mendex, and Schaposnik}}]{anom1}
\bibinfo{author}{\bibfnamefont{P.}~\bibnamefont{Arias}},
  \bibinfo{author}{\bibnamefont{H.Falomir}},
  \bibinfo{author}{\bibfnamefont{J.}~\bibnamefont{Gamboa}},
  \bibinfo{author}{\bibfnamefont{F.}~\bibnamefont{Mendex}}, \bibnamefont{and}
  \bibinfo{author}{\bibfnamefont{F.}~\bibnamefont{Schaposnik}},
  \bibinfo{journal}{Phys.\ Rev.\ D} \textbf{\bibinfo{volume}{76}},
  \bibinfo{pages}{025019} (\bibinfo{year}{2007}).

\bibitem[{\citenamefont{Salvio}(2008)}]{anom2}
\bibinfo{author}{\bibfnamefont{A.}~\bibnamefont{Salvio}},
  \bibinfo{journal}{Phys.\ Rev.\ D} \textbf{\bibinfo{volume}{78}},
  \bibinfo{pages}{085023} (\bibinfo{year}{2008}).

\bibitem[{\citenamefont{Colladay and McDonald}(2007)}]{cm1}
\bibinfo{author}{\bibfnamefont{D.}~\bibnamefont{Colladay}} \bibnamefont{and}
  \bibinfo{author}{\bibfnamefont{P.}~\bibnamefont{McDonald}},
  \bibinfo{journal}{Phys.\ Rev.\ D} \textbf{\bibinfo{volume}{75}},
  \bibinfo{pages}{105002} (\bibinfo{year}{2007}).

\bibitem[{\citenamefont{Colladay and McDonald}(2008)}]{cm2}
\bibinfo{author}{\bibfnamefont{D.}~\bibnamefont{Colladay}} \bibnamefont{and}
  \bibinfo{author}{\bibfnamefont{P.}~\bibnamefont{McDonald}},
  \bibinfo{journal}{Phys.\ Rev.\ D} \textbf{\bibinfo{volume}{77}},
  \bibinfo{pages}{085006} (\bibinfo{year}{2008}).

\bibitem[{\citenamefont{Kosteleck\'y and Samuel}(1989{\natexlab{a}})}]{kps1}
\bibinfo{author}{\bibfnamefont{V.}~\bibnamefont{Kosteleck\'y}}
  \bibnamefont{and} \bibinfo{author}{\bibfnamefont{S.}~\bibnamefont{Samuel}},
  \bibinfo{journal}{Phys.\ Rev.\ D} \textbf{\bibinfo{volume}{39}},
  \bibinfo{pages}{683} (\bibinfo{year}{1989}{\natexlab{a}}).

\bibitem[{\citenamefont{Kosteleck\'y and Samuel}(1989{\natexlab{b}})}]{kps2}
\bibinfo{author}{\bibfnamefont{V.}~\bibnamefont{Kosteleck\'y}}
  \bibnamefont{and} \bibinfo{author}{\bibfnamefont{S.}~\bibnamefont{Samuel}},
  \bibinfo{journal}{Phys.\ Rev.\ D} \textbf{\bibinfo{volume}{40}},
  \bibinfo{pages}{1886} (\bibinfo{year}{1989}{\natexlab{b}}).

\bibitem[{\citenamefont{Kosteleck\'y and Samuel}(1989{\natexlab{c}})}]{kps3}
\bibinfo{author}{\bibfnamefont{V.}~\bibnamefont{Kosteleck\'y}}
  \bibnamefont{and} \bibinfo{author}{\bibfnamefont{S.}~\bibnamefont{Samuel}},
  \bibinfo{journal}{Phys.\ Rev.\ Lett.} \textbf{\bibinfo{volume}{63}},
  \bibinfo{pages}{224} (\bibinfo{year}{1989}{\natexlab{c}}).

\bibitem[{\citenamefont{Kosteleck\'y and Samuel}(1991)}]{kps4}
\bibinfo{author}{\bibfnamefont{V.}~\bibnamefont{Kosteleck\'y}}
  \bibnamefont{and} \bibinfo{author}{\bibfnamefont{S.}~\bibnamefont{Samuel}},
  \bibinfo{journal}{Phys.\ Rev.\ D} \textbf{\bibinfo{volume}{66}},
  \bibinfo{pages}{1811} (\bibinfo{year}{1991}).

\bibitem[{\citenamefont{Kosteleck\'y and Samuel}(1989{\natexlab{d}})}]{kps5}
\bibinfo{author}{\bibfnamefont{V.}~\bibnamefont{Kosteleck\'y}}
  \bibnamefont{and} \bibinfo{author}{\bibfnamefont{S.}~\bibnamefont{Samuel}},
  \bibinfo{journal}{Phys.\ Rev.\ D} \textbf{\bibinfo{volume}{39}},
  \bibinfo{pages}{683} (\bibinfo{year}{1989}{\natexlab{d}}).

\bibitem[{\citenamefont{Kosteleck\'y and Potting}(1991)}]{kps6}
\bibinfo{author}{\bibfnamefont{V.}~\bibnamefont{Kosteleck\'y}}
  \bibnamefont{and} \bibinfo{author}{\bibfnamefont{R.}~\bibnamefont{Potting}},
  \bibinfo{journal}{Nucl.\ Phys.\ B} \textbf{\bibinfo{volume}{359}},
  \bibinfo{pages}{545} (\bibinfo{year}{1991}).

\bibitem[{\citenamefont{Kosteleck\'y and Potting}(1996)}]{kps7}
\bibinfo{author}{\bibfnamefont{V.}~\bibnamefont{Kosteleck\'y}}
  \bibnamefont{and} \bibinfo{author}{\bibfnamefont{R.}~\bibnamefont{Potting}},
  \bibinfo{journal}{Phys.\ Lett.\ B} \textbf{\bibinfo{volume}{381}},
  \bibinfo{pages}{89} (\bibinfo{year}{1996}).

\bibitem[{\citenamefont{Kosteleck\'y and Potting}(2001)}]{kps8}
\bibinfo{author}{\bibfnamefont{V.}~\bibnamefont{Kosteleck\'y}}
  \bibnamefont{and} \bibinfo{author}{\bibfnamefont{R.}~\bibnamefont{Potting}},
  \bibinfo{journal}{Phys.\ Rev.\ D} \textbf{\bibinfo{volume}{63}},
  \bibinfo{pages}{046007} (\bibinfo{year}{2001}).

\bibitem[{\citenamefont{Kosteleck\'y et~al.}(2000)\citenamefont{Kosteleck\'y,
  Perry, and Potting}}]{kps9}
\bibinfo{author}{\bibfnamefont{V.}~\bibnamefont{Kosteleck\'y}},
  \bibinfo{author}{\bibfnamefont{M.}~\bibnamefont{Perry}}, \bibnamefont{and}
  \bibinfo{author}{\bibfnamefont{R.}~\bibnamefont{Potting}},
  \bibinfo{journal}{Phys.\ Rev.\ Lett.} \textbf{\bibinfo{volume}{84}},
  \bibinfo{pages}{4541} (\bibinfo{year}{2000}).

\bibitem[{\citenamefont{Kosteleck\'y and Mewes}(2007)}]{kmewes}
\bibinfo{author}{\bibfnamefont{V.}~\bibnamefont{Kosteleck\'y}}
  \bibnamefont{and} \bibinfo{author}{\bibfnamefont{M.}~\bibnamefont{Mewes}},
  \bibinfo{journal}{Phys.\ Rev.\ Lett.} \textbf{\bibinfo{volume}{99}},
  \bibinfo{pages}{011601} (\bibinfo{year}{2007}).

\bibitem[{\citenamefont{Kostelecky}(2004)}]{alangrav1}
\bibinfo{author}{\bibfnamefont{V.}~\bibnamefont{Kostelecky}},
  \bibinfo{journal}{Phys.\ Rev.\ D} \textbf{\bibinfo{volume}{69}},
  \bibinfo{pages}{105009} (\bibinfo{year}{2004}).

\bibitem[{\citenamefont{Bluhm and Kostelecky}(2005)}]{alangrav2}
\bibinfo{author}{\bibfnamefont{R.}~\bibnamefont{Bluhm}} \bibnamefont{and}
  \bibinfo{author}{\bibfnamefont{V.}~\bibnamefont{Kostelecky}},
  \bibinfo{journal}{Phys.\ Rev.\ D} \textbf{\bibinfo{volume}{71}},
  \bibinfo{pages}{065008} (\bibinfo{year}{2005}).

\bibitem[{\citenamefont{Bailey and Kostelecky}(2006)}]{alangrav3}
\bibinfo{author}{\bibfnamefont{Q.}~\bibnamefont{Bailey}} \bibnamefont{and}
  \bibinfo{author}{\bibfnamefont{V.}~\bibnamefont{Kostelecky}},
  \bibinfo{journal}{Phys.\ Rev.\ D} \textbf{\bibinfo{volume}{74}},
  \bibinfo{pages}{045001} (\bibinfo{year}{2006}).

\bibitem[{\citenamefont{Kostelecky and Potting}(2005)}]{kpa}
\bibinfo{author}{\bibfnamefont{V.}~\bibnamefont{Kostelecky}} \bibnamefont{and}
  \bibinfo{author}{\bibfnamefont{R.}~\bibnamefont{Potting}},
  \bibinfo{journal}{Gen.\ Rel.\ Grav.} \textbf{\bibinfo{volume}{37}},
  \bibinfo{pages}{1675} (\bibinfo{year}{2005}).

\bibitem[{\citenamefont{Kostelecky and Potting}()}]{kpb}
\bibinfo{author}{\bibfnamefont{V.}~\bibnamefont{Kostelecky}} \bibnamefont{and}
  \bibinfo{author}{\bibfnamefont{R.}~\bibnamefont{Potting}},
  \bibinfo{note}{arXiv:0901.0662.}

\bibitem[{\citenamefont{Colladay and McDonald}(2002)}]{cmjmp1}
\bibinfo{author}{\bibfnamefont{D.}~\bibnamefont{Colladay}} \bibnamefont{and}
  \bibinfo{author}{\bibfnamefont{P.}~\bibnamefont{McDonald}},
  \bibinfo{journal}{J.\ Math.\ Phys.} \textbf{\bibinfo{volume}{43}},
  \bibinfo{pages}{3554} (\bibinfo{year}{2002}).

\bibitem[{\citenamefont{Colladay and McDonald}(2004)}]{cmjmp2}
\bibinfo{author}{\bibfnamefont{D.}~\bibnamefont{Colladay}} \bibnamefont{and}
  \bibinfo{author}{\bibfnamefont{P.}~\bibnamefont{McDonald}},
  \bibinfo{journal}{J.\ Math.\ Phys.} \textbf{\bibinfo{volume}{45}},
  \bibinfo{pages}{3228} (\bibinfo{year}{2004}).

\bibitem[{\citenamefont{Frank and Turan}(2006)}]{ft}
\bibinfo{author}{\bibfnamefont{M.}~\bibnamefont{Frank}} \bibnamefont{and}
  \bibinfo{author}{\bibfnamefont{I.}~\bibnamefont{Turan}},
  \bibinfo{journal}{Phys.\ Rev.\ D} \textbf{\bibinfo{volume}{74}},
  \bibinfo{pages}{033016} (\bibinfo{year}{2006}).

\bibitem[{\citenamefont{Altschul}(2004{\natexlab{c}})}]{ba}
\bibinfo{author}{\bibfnamefont{B.}~\bibnamefont{Altschul}},
  \bibinfo{journal}{Phys.\ Rev.\ D} \textbf{\bibinfo{volume}{70}},
  \bibinfo{pages}{101701} (\bibinfo{year}{2004}{\natexlab{c}}).

\bibitem[{\citenamefont{Altschul and Kostelecky}(2005)}]{kalt}
\bibinfo{author}{\bibfnamefont{B.}~\bibnamefont{Altschul}} \bibnamefont{and}
  \bibinfo{author}{\bibfnamefont{A.}~\bibnamefont{Kostelecky}},
  \bibinfo{journal}{Phys.\ Lett.\ B} \textbf{\bibinfo{volume}{628}},
  \bibinfo{pages}{106} (\bibinfo{year}{2005}).

\bibitem[{\citenamefont{Altschul}(2006)}]{altghost}
\bibinfo{author}{\bibfnamefont{B.}~\bibnamefont{Altschul}},
  \bibinfo{journal}{Phys.\ Rev.\ D} \textbf{\bibinfo{volume}{73}},
  \bibinfo{pages}{045004} (\bibinfo{year}{2006}).

\bibitem[{\citenamefont{Anselmi and Halat}(2007)}]{ans1}
\bibinfo{author}{\bibfnamefont{D.}~\bibnamefont{Anselmi}} \bibnamefont{and}
  \bibinfo{author}{\bibfnamefont{M.}~\bibnamefont{Halat}},
  \bibinfo{journal}{Phys.\ Rev.\ D} \textbf{\bibinfo{volume}{76}},
  \bibinfo{pages}{125011} (\bibinfo{year}{2007}).

\bibitem[{\citenamefont{Anselmi}(2008)}]{ans2}
\bibinfo{author}{\bibfnamefont{D.}~\bibnamefont{Anselmi}}
  (\bibinfo{year}{2008}), \bibinfo{note}{arXiv:0808.3470; arXiv:0808.3474}.

\bibitem[{\citenamefont{Mariz et~al.}()\citenamefont{Mariz, Nascimento, Petrov,
  Santos, and de~Silva}}]{funcdet}
\bibinfo{author}{\bibfnamefont{T.}~\bibnamefont{Mariz}},
  \bibinfo{author}{\bibfnamefont{J.~R.} \bibnamefont{Nascimento}},
  \bibinfo{author}{\bibfnamefont{A.~Y.} \bibnamefont{Petrov}},
  \bibinfo{author}{\bibfnamefont{L.~Y.} \bibnamefont{Santos}},
  \bibnamefont{and} \bibinfo{author}{\bibfnamefont{A.~J.}
  \bibnamefont{de~Silva}}, \bibinfo{note}{arXiv:0708.3348}.

\bibitem[{\citenamefont{Colladay and Kostelecky}(1997)}]{ck1}
\bibinfo{author}{\bibfnamefont{D.}~\bibnamefont{Colladay}} \bibnamefont{and}
  \bibinfo{author}{\bibfnamefont{V.~A.} \bibnamefont{Kostelecky}},
  \bibinfo{journal}{Phys.\ Rev.\ D} \textbf{\bibinfo{volume}{55}},
  \bibinfo{pages}{6760} (\bibinfo{year}{1997}).

\bibitem[{\citenamefont{Colladay and Kostelecky}(1998)}]{ck2}
\bibinfo{author}{\bibfnamefont{D.}~\bibnamefont{Colladay}} \bibnamefont{and}
  \bibinfo{author}{\bibfnamefont{V.~A.} \bibnamefont{Kostelecky}},
  \bibinfo{journal}{Phys.\ Rev.\ D} \textbf{\bibinfo{volume}{58}},
  \bibinfo{pages}{116002} (\bibinfo{year}{1998}).

\bibitem[{\citenamefont{Peskin and D.Schroeder}(1995)}]{ps}
\bibinfo{author}{\bibfnamefont{M.}~\bibnamefont{Peskin}} \bibnamefont{and}
  \bibinfo{author}{\bibnamefont{D.Schroeder}}, \emph{\bibinfo{title}{An
  Introduction to Quantum Field Theory}} (\bibinfo{publisher}{Perseus Books},
  \bibinfo{year}{1995}).

\bibitem[{\citenamefont{Itzykson and Zuber}(1980)}]{iz}
\bibinfo{author}{\bibfnamefont{C.}~\bibnamefont{Itzykson}} \bibnamefont{and}
  \bibinfo{author}{\bibfnamefont{J.}~\bibnamefont{Zuber}},
  \emph{\bibinfo{title}{Quantum Field Theory}}
  (\bibinfo{publisher}{McGraw-Hill}, \bibinfo{year}{1980}).

\end{thebibliography}

\end{document}